\documentclass[twoside,english]{elsarticle}
\usepackage[T1]{fontenc}
\usepackage[latin9]{inputenc}
\usepackage{amsbsy}
\usepackage{amstext}
\usepackage{esint}

\makeatletter
\journal{Example: Physics Letters B}

\makeatother

\usepackage{babel}
\begin{document}

\title{Perturbative quantum damping of cosmological expansion}

\author{Bogus\l{}aw~Broda}

\ead{bobroda@uni.lodz.pl}

\address{Department of Theoretical Physics, Faculty of Physics and Applied
Informatics, University of \L{}ód\'{z}, Pomorska 149/153, 90-236 \L{}ód\'{z},
Poland}
\begin{abstract}
Perturbative quantum gravity in the framework of the Schwinger--Keldysh
formalism is applied to compute lowest-order corrections to the actual
expansion of the Universe described in terms of the spatially flat
Friedman--Lema{\^i}tre--Robertson--Walker solution. The classical
metric is approximated by a third order polynomial perturbation around
the Minkowski metric. It is shown that the quantum contribution to
the classical expansion, although extremely small, has damping properties
(quantum friction), i.e.\ it slows down the expansion.\end{abstract}
\begin{keyword}
perturbative quantum gravity \sep cosmological expansion \sep Schwinger--Keldysh
formalism \PACS 04.60.Gw \sep 04.60.Bc \sep 04.60.Pp \sep 98.80.Qc
\end{keyword}
\maketitle

\section{Introduction}

The aim of our work is to explicitly show the appearance of quantum
generated damping or quantum friction, i.e. slowing down, in the present
(accelerating) expansion of the Universe. In general, quantum corrections
to classical gravitational field can be perturbatively calculated
in a number of ways. First of all, it is possible to directly calculate
quantum (one-loop) corrections to classical gravitational field from
the graviton vacuum polarization (self-energy), in the analogy to
the case of the Coulomb potential in QED (for example, see \citet{QuantumElectrodynamics}),
the so-called Uehling potential. Such a type of calculations has been
already performed for the Schwarzschild solution by \citet{duff1974quantum},
as well as for the spatially flat Friedman--Lema{\^i}tre--Robertson--Walker
(FLRW) metric (\citet{broda2011one}). Another approach refers to
the energy-momentum tensor calculations, and it has been applied to
the Newton potential (for example, see \citet{bjerrum2003quantum},
and the references therein), to the Reissner-Nordstr\"om and the
Kerr-Newman solutions (see \citet{donoghue2002quantum}), and to the
Schwarzschild and the Kerr solution metrics (see \citet{jannik2002quantum}).
Yet another approach uses the Schwinger--Keldysh (SK) formalism to
the case of the Newton potential (e.g., see \citet{park2010solving}).
It is argued that only the SK formalism is adequate for time-dependent
potentials, hence in particular, in the context of cosmology (e.g.,
see \citet{weinberg2005quantum}, and references therein). As we aim
to perturbatively calculate corrections to the spatially flat FLRW
metric, we should use the SK formalism, because this is exactly that
case (time-dependence of gravitational field) the SK approach has
been devised for.

The corrections we calculate are a quantum response to the spatially
flat FLRW solution which is described by a small perturbation around
the Minkowski metric. Since such a type of calculations is usually
plagued by infinities, we confine ourselves to the classical perturbation
given by a polynomial of the third degree. Moreover, to avoid infinities
on intermediate stages of our calculations, time derivative of the
convolution of the time propagator with the perturbation of the metric
should be performed in a suitable order. The final result is given
in terms of the present time quantum correction $q_{0}^{\textrm{Q}}$
to the deceleration parameter $q_{0}$. Interestingly, it appears
that $q_{0}^{\textrm{Q}}$ is positive, although obviously, it is
extremely small.

\section{Quantum damping}

Our starting point is a general spatially flat FLRW metric

\begin{equation}
ds^{2}\equiv g_{\mu\nu}dx^{\mu}dx^{\nu}=-dt^{2}+a^{2}\left(t\right)d\mathbf{\boldsymbol{x}}{}^{2},\qquad\mu,\nu=0,1,2,3,
\end{equation}
with the cosmological scale factor $a\left(t\right)$. To satisfy
the condition of weakness of the perturbative gravitational field
$h_{\mu\nu}$ near our reference time $t=t_{0}$ (where $t_{0}$ could
be the age of the Universe --- the present moment) in the expansion
\begin{equation}
g_{\mu\nu}\left(x\right)=\eta_{\mu\nu}+h_{\mu\nu}\left(x\right),
\end{equation}
the metric should be normalized in such a way that it is exactly Minkowskian
for $t=t_{0}$, i.e.
\begin{equation}
a^{2}\left(t\right)=1+h\left(t\right),\qquad h\left(t_{0}\right)=0.
\end{equation}
Let us note the analogy to the Newton potential ($\sim1/r$), where
the ``reference radius'' is in spatial infinity, i.e.\ $r_{0}=+\infty$.
Then, in the block diagonal form,
\begin{equation}
h_{\mu\nu}\left(t,\boldsymbol{x}\right)=\left(\begin{array}{cc}
0 & 0\\
0 & \delta_{ij}h\left(t\right)
\end{array}\right),\quad i,j=1,2,3.\label{eq:hmini}
\end{equation}

To obtain quantum corrections to the classical gravitational field
$h_{\mu\nu}^{\textrm{C }}\left(x\right)$, we shall use the one-loop
effective field equation derived by \citet{park2010solving} 
\begin{eqnarray}
\mathcal{D}^{\mu\nu\varrho\sigma}h_{\varrho\sigma}^{\textrm{Q}}\left(t,\boldsymbol{x}\right) & = & \frac{\kappa^{2}}{10240\pi^{3}}\textrm{D}^{\mu\nu\varrho\sigma}\partial^{4}\intop_{0}^{t}dt'\nonumber \\
 & \times & \int d^{3}x'\theta\left(\Delta t-\Delta r\right)\left[\ln\left(-\mu^{2}\Delta x^{2}\right)-1\right]h_{\varrho\sigma}^{\textrm{C }}\left(t',\boldsymbol{x}'\right),\label{eq:effeq}
\end{eqnarray}
where $\Delta t\equiv t-t'$, $\Delta r\equiv\left|\boldsymbol{x}-\boldsymbol{x}'\right|$,
$\Delta x^{2}\equiv-\left(\Delta t\right)^{2}+\left(\Delta r\right)^{2}$,
and the mass scale $\mu$ is coming from the renormalization procedure
(see \citet{ford2005stress}). Here $\kappa^{2}=16\pi G_{N}$, where
$G_{N}$ is the Newton gravitational constant. The operator $\mathcal{D}$
(the Lichnerowicz operator in the flat background) is of the form
\[
\mathcal{D}^{\mu\nu\varrho\sigma}=\frac{1}{2}\left(\eta^{\mu\nu}\eta^{\varrho\sigma}\partial^{2}-\partial^{\mu}\partial^{\nu}\eta^{\varrho\sigma}-\eta^{\mu\nu}\partial^{\varrho}\partial^{\sigma}-\eta^{\mu(\varrho}\eta^{\sigma)\nu}\partial^{2}+2\partial^{(\mu}\eta^{\nu)(\varrho}\partial^{\sigma)}\right),
\]
whereas for the minimally coupled massless scalar field
\begin{eqnarray}
\textrm{D}{}^{\mu\nu\varrho\sigma} & = & \Pi^{\mu\nu}\Pi^{\varrho\sigma}+\frac{1}{3}\Pi^{\mu(\varrho}\Pi^{\sigma)\nu}\label{eq:opd}
\end{eqnarray}
with
\[
\Pi^{\mu\nu}\equiv\eta^{\mu\nu}\partial^{2}-\partial^{\mu}\partial^{\nu}.
\]
For conformally coupled fields we have $\widetilde{\textrm{D}}$ instead
of $\textrm{D}$, where
\[
\widetilde{\textrm{D}}{}^{\mu\nu\varrho\sigma}\equiv-\frac{1}{9}\Pi^{\mu\nu}\Pi^{\varrho\sigma}+\frac{1}{3}\Pi^{\mu(\varrho}\Pi^{\sigma)\nu}.
\]
Since the metric depends only on time, we can explicitly perform
the spatial integration in (\ref{eq:effeq}) with respect to $\boldsymbol{x}'$
obtaining the integral kernel (time propagator)
\begin{eqnarray}
\textrm{K}\left(\Delta t\right) & \equiv & 4\pi\intop_{0}^{\Delta t}dr\, r^{2}\left\{ \ln\left[\mu^{2}\left(\left(\Delta t\right)^{2}-r^{2}\right)\right]-1\right\} \nonumber \\
 & = & \frac{4\pi}{3}\left(\Delta t\right)^{3}\left[\ln\left(4\mu^{2}\Delta t^{2}\right)-\frac{11}{3}\right].\label{eq:defk}
\end{eqnarray}
For the time-dependent metric of the form
\[
\left(\begin{array}{cc}
f\left(t\right)\\
 & \delta_{ij}h\left(t\right)
\end{array}\right),
\]
the action of the operators $\mathcal{D}$, $\textrm{D}$ and $\widetilde{\textrm{D}}$
is given by
\begin{equation}
\mathcal{D}\left(\begin{array}{cc}
f\left(t\right)\\
 & \delta_{ij}h\left(t\right)
\end{array}\right)=\left(\begin{array}{cc}
0\\
 & -\delta_{ij}\frac{d^{2}}{dt^{2}}h\left(t\right)
\end{array}\right),\label{eq:dcalonmet}
\end{equation}
\begin{eqnarray}
\textrm{D}\left(\begin{array}{cc}
f\left(t\right) & 0\\
0 & \delta_{ij}h\left(t\right)
\end{array}\right) & = & \left(\begin{array}{cc}
0 & 0\\
0 & \frac{10}{3}\delta_{ij}\frac{d^{4}}{dt^{4}}h\left(t\right)
\end{array}\right),\label{eq:dromonmet}
\end{eqnarray}
whereas
\begin{equation}
\widetilde{\textrm{D}}\left(\begin{array}{cc}
a\left(t\right) & 0\\
0 & \delta_{ij}h\left(t\right)
\end{array}\right)=0,\label{eq:dtilonmet}
\end{equation}
respectively. There are no mixing of diagonal and non-diagonal terms,
and the empty blocks mean expressions which can be non-zero, but they
are inessential in our further analysis. Thus, (\ref{eq:effeq}) assumes
the simple form
\begin{equation}
\frac{d^{2}}{dt^{2}}h^{\textrm{Q}}\left(t\right)=-\frac{\kappa^{2}}{3072\pi^{3}}\frac{d^{8}}{dt^{8}}\left(\textrm{K}\star h^{\textrm{C}}\right)\left(t\right),\label{eq:diffeq}
\end{equation}
where the integral kernel $\textrm{K}$ is given in (\ref{eq:defk}),
and the convolution ``$\star$'' is standardly defined by
\begin{equation}
\left(\textrm{K}\star\mathrm{F}\right)\left(t\right)\equiv\intop_{0}^{t}\textrm{K}\left(t-t'\right)\mathrm{F}\left(t'\right)dt'=\intop_{0}^{t}\textrm{K}\left(t'\right)\mathrm{F}\left(t-t'\right)dt'.\label{eq:defconv}
\end{equation}
One should note that due to ``diagonality'' of (\ref{eq:hmini})
and (\ref{eq:dcalonmet}--\ref{eq:dtilonmet}), no non-diagonal terms
of the metric enter (\ref{eq:diffeq}).

Now, observing that also the limit of integration in (\ref{eq:defconv})
depends on $t$, one can easily derive the following differentiation
formula for the convolution
\begin{equation}
\frac{d^{n}}{dt^{n}}\,\left(\textrm{K}\star\mathrm{F}\right)\left(t\right)=\left(\frac{d^{n}}{dt^{n}}\textrm{K}\star\mathrm{F}\right)\left(t\right)+{\displaystyle \sum_{k=1}^{n}\frac{d^{\left(n-k\right)}}{dt^{\left(n-k\right)}}\textrm{K}\left(0\right)\frac{d^{\left(k-1\right)}}{dt^{\left(k-1\right)}}\textrm{F}\left(t\right).}\label{eq:diffofconv}
\end{equation}
Using symmetry between $\textrm{K}$ and $\textrm{F}$ (see (\ref{eq:defconv})),
it is possible to distribute differentiation in (\ref{eq:diffofconv})
in several different ways. For practical purposes, i.e.\ elimination
of possible singular terms on intermediate stages of our calculations,
the most convenient form of the eighth derivative is the symmetric
one, i.e.
\begin{eqnarray}
\frac{d^{8}}{dt^{8}}\left(\textrm{K}\star h^{\textrm{C}}\right)\left(t\right) & = & \left(\frac{d^{4}}{dt^{4}}\textrm{K}\star\frac{d^{4}}{dt^{4}}h^{\textrm{C}}\right)\left(t\right)\nonumber \\
 & + & \sum_{k=1}^{4}\left[\frac{d^{\left(4-k\right)}}{dt^{\left(4-k\right)}}\textrm{K}\left(0\right)\frac{d^{\left(k+3\right)}}{dt^{\left(k+3\right)}}h^{\mathrm{C}}\left(t\right)\right.\label{eq:mixdiff}\\
 & + & \left.\frac{d^{\left(4-k\right)}}{dt^{\left(4-k\right)}}h^{\mathrm{C}}\left(0\right)\frac{d^{\left(k+3\right)}}{dt^{\left(k+3\right)}}\textrm{K}\left(t\right)\right].\nonumber 
\end{eqnarray}
To finally prevent the appearance of possible infinities, i.e.\ primary
UV infinities in the propagator, signaled by $\mu$, as well as divergences
in the convolution, which could come from singularities in the kernel
(time propagator) $\textrm{K}$, we assume the following third order
polynomial form of the classical metric
\begin{equation}
h^{\textrm{C}}\left(\tau\right)=h_{0}+h_{1}\tau+h_{2}\tau^{2}+h_{3}\tau^{3}.\label{eq:polh}
\end{equation}
Henceforth, for simplicity, we use the dimensionless unit of time,
$\tau\equiv t/t_{0}$, instead of $t$.

Non-singularity of Eq.\ref{eq:eightdiff} proofs that (\ref{eq:mixdiff})
and (\ref{eq:polh}), have been properly selected. In fact, our choice
is unique. First of all, let us observe that
\[
\left.\frac{d^{k}}{dt^{k}}\textrm{K}\left(t\right)\right|_{t=0},\qquad\textrm{for}\quad k>2,
\]
is singular, and
\[
\frac{d^{k}}{dt^{k}}\textrm{K}\left(t\right),\qquad\textrm{for}\quad k<4,
\]
is $\mu$-dependent. Then, the only possibility to avoid such troublesome
terms admits exactly the products in the second part of the sum in
(\ref{eq:mixdiff}). In turn, to nullify the unwanted first part of
the sum in (\ref{eq:mixdiff}), $h^{\textrm{C}}\left(\tau\right)$
should be of the form (\ref{eq:polh}). The term before the summation
sign in (\ref{eq:mixdiff}) vanishes, and thus, it is inessential.

Actually, the classical metric (\ref{eq:polh}) does not belong to
any favorite family of cosmological solutions, perhaps except for
the linear case ($h_{0}=-1,h_{1}=1,h_{2}=h_{3}=0$), corresponding
to radiation. In fact, a physically realistic metric is not precisely
given, for example, by the matter-dominated cosmological scale factor
$a\left(\tau\right)=\tau^{2/3}$, because firstly, the character of
cosmological evolution depends on the epoch (time $\tau$), and secondly,
it is ``contaminated'' by other ``matter'' components, e.g.\ radiation,
and possibly, dark energy. Therefore, we should consider (\ref{eq:polh})
as a phenomenological description, approximating actual cosmological
evolution on the finite time interval $\tau\in\left[\tau_{0},1\right]$,
$0\leq\tau_{0}<1$.

Inserting (\ref{eq:defk}) and (\ref{eq:polh}) to (\ref{eq:mixdiff})
we derive, by virtue of (\ref{eq:diffeq}), the second order differential
equation
\begin{equation}
\ddot{h}^{\textrm{Q}}\left(\tau\right)=\lambda\left(h_{0}\tau^{-2}-\frac{h_{1}}{3}\tau^{-1}+\frac{h_{2}}{3}-h_{3}\tau\right),\label{eq:eightdiff}
\end{equation}
which can be easily integrated out with respect to $\tau$, yielding
\begin{equation}
\dot{h}^{\textrm{Q}}\left(\tau\right)=\lambda\left(-h_{0}\tau^{-1}-\frac{h_{1}}{3}\log\left|\tau\right|+\frac{h_{2}}{3}\tau-\frac{h_{3}}{2}\tau^{2}\right)\label{eq:seventhdiff}
\end{equation}
and
\begin{equation}
h^{\textrm{Q}}=\lambda\left(-h_{0}\log\left|\tau\right|-\frac{h_{1}}{3}\left(\tau\log\left|\tau\right|-\tau\right)+\frac{h_{2}}{6}\tau^{2}-\frac{h_{3}}{6}\tau^{3}\right),\label{eq:sixthdiff}
\end{equation}
where $\lambda\equiv\kappa^{2}/32\pi^{2}t_{0}^{2}\approx\frac{1}{2}\cdot10^{-46}$.
As a physical observable we are interested in, we take the deceleration
parameter
\begin{equation}
q\left(\tau\right)\equiv-\frac{a\left(\tau\right)\ddot{a}\left(\tau\right)}{\dot{a}^{2}\left(\tau\right)}=1-2\left[1+h\left(\tau\right)\right]\frac{\ddot{h}\left(\tau\right)}{\dot{h}^{2}\left(\tau\right)}.\label{eq:qzero-1}
\end{equation}
The quantum contribution to the deceleration parameter, namely, the
lowest order contribution of (\ref{eq:eightdiff}--\ref{eq:sixthdiff})
to (\ref{eq:qzero-1}), i.e.\ $q\left(\tau\right)=q^{\textrm{C}}\left(\tau\right)+q^{\textrm{Q}}\left(\tau\right)+\mathcal{O}\left(\lambda^{2}\right)$,
reads
\begin{equation}
q^{\textrm{Q}}=-\frac{2}{\left(\dot{h}^{\textrm{C}}\right)^{2}}\left[\ddot{h}^{\textrm{C}}h^{\textrm{Q}}+\left(1+h^{\textrm{C}}\right)\left(\ddot{h}^{\textrm{Q}}-\frac{2\ddot{h}^{\textrm{C}}\dot{h}^{\textrm{Q}}}{\dot{h}^{\textrm{C}}}\right)\right].\label{eq:qquant}
\end{equation}
To approximate the cosmological evolution by the (four-parameter)
phenomenological metric (\ref{eq:polh}), we need four conditions.
First of all, we impose the following two obvious boundary conditions
\[
h^{\textrm{C}}\left(0\right)=-1\quad\textrm{and}\quad h^{\textrm{C}}\left(1\right)=0,
\]
corresponding to
\[
a^{2}\left(0\right)=0\quad\textrm{and}\quad a^{2}\left(1\right)=1,
\]
and implying
\begin{equation}
h_{0}=-1\quad\textrm{and}\quad h_{1}+h_{2}+h_{3}=-h_{0}=1.\label{eq:firsttwoeq}
\end{equation}
In this place various different further directions of proceedings
could be assumed, depending on the question we pose.

Then, let us study the quantum contribution to the actual cosmological
evolution. By virtue of (\ref{eq:defk}) and (\ref{eq:diffeq}), the
``effective'' time propagator determined by the sixth order derivative
of the kernel $\textrm{K}$, behaves as $\left(\Delta t\right)^{-3}$,
which follows from, e.g., dimensional analysis. Thus, the largest
contributions to the quantum part of the metric $h^{\textrm{Q}}\left(\tau\right)$
are coming from integration (\ref{eq:defconv}) in the vicinity of
$\tau\approx1$, because of the large value of $\left(\tau-1\right)^{-3}$.
Therefore, we impose the next two additional conditions at the dominating
point $\tau=1$. Namely, $h^{\textrm{C}}$ is supposed to yield the
observed value of the Hubble constant
\begin{equation}
H_{0}\equiv\frac{\dot{a}\left(1\right)}{a\left(1\right)}=\frac{1}{2}\dot{h}^{\textrm{C}}\left(1\right),\label{eq:hubbledef}
\end{equation}
and the observed deceleration parameter $q_{0}=q^{\textrm{C}}\left(1\right)$.
Solving (\ref{eq:qzero-1}), (\ref{eq:firsttwoeq}) and (\ref{eq:hubbledef})
for $h_{k}$ ($k=1,2,3$), we obtain
\begin{eqnarray}
h_{1} & = & 3-\left(3+q_{0}\right)H_{0},\nonumber \\
h_{2} & = & -3+\left(4+2q_{0}\right)H_{0},\label{eq:threeh}\\
h_{3} & = & 1-\left(1+q_{0}\right)H_{0}.\nonumber 
\end{eqnarray}
To estimate only the order and the qualitative behavior of the present
time quantum contribution to the accelerating expansion of the Universe,
it is sufficient to insert to (\ref{eq:threeh}) the following crude
approximation: $H_{0}=1$ and $q_{0}=-\frac{1}{2}$. Now
\[
h_{1}=\frac{1}{2},\qquad h_{2}=0,\qquad h_{3}=\frac{1}{2},
\]
yielding 
\[
\dot{h}^{\textrm{C}}\left(1\right)=2,\qquad\ddot{h}^{\textrm{C}}\left(1\right)=3.
\]
Finally, by virtue of (\ref{eq:eightdiff}--\ref{eq:sixthdiff})
\[
h^{\textrm{Q}}\left(1\right)=\frac{\lambda}{12},\qquad\dot{h}^{\textrm{Q}}\left(1\right)=\frac{3\lambda}{4},\qquad\ddot{h}^{\textrm{C}}\left(1\right)=-\frac{5\lambda}{3},
\]
and hence (see (\ref{eq:qquant}))
\begin{equation}
q_{0}^{\textrm{Q}}\equiv q^{\textrm{Q}}\left(1\right)=\frac{11\lambda}{6}=\frac{11\kappa^{2}}{192\pi^{2}t_{0}^{2}}\sim10^{-46}.\label{eq:finalnumber}
\end{equation}

\section{Summary}

In the framework of the SK (one-loop) perturbative quantum gravity,
we have derived the formula (\ref{eq:finalnumber}) expressing the
(approximated) value of the present time quantum contribution $q_{0}^{\textrm{Q}}$
to the deceleration parameter $q_{0}$. The present time quantum contribution,
$q_{0}^{\textrm{Q}}\sim+10^{-46}$, is positive but it is negligibly
small in comparison to the observed (negative) value of the deceleration
parameter, $q_{0}\approx-\frac{1}{2}$. Therefore, we deal with an
extremely small damping (slowing down) of the expansion of the Universe,
which is of quantum origin (quantum friction).

One should also stress, that in the course of our analysis, because
of some technical difficulties, we have been forced to confine our
work to a particular case: a conforming with the FLRW form polynomial
(of the third degree) perturbation around the Minkowski metric ---
to avoid short distance infinities; minimally coupled massless scalar
field and conformally coupled fields (trivial contributions) only,
without (virtual) gravitons --- to avoid calculational limitations.

Finally, it would be desirable to compare our present result to our
earlier computation (see \citet{broda2011one}), where we have obtained
an opposite result, i.e.\ repulsion instead of damping. First of
all, one should note that non-SK approaches are acausal, in general,
for finite time intervals, as they take into account contributions
coming from the future state of the Universe. This follows from the
fact that the Feynman propagator has an ``advanced tail'', which
is not dangerous for (infinite time interval) S-matrix elements. Besides,
the present work concerns scalar field contributions, whereas the
result of the previous paper is determined by graviton contributions.
Nevertheless, in the both approaches, quantum contributions are trivial
for conformal fields, which well corresponds to conformal flatness
of the FLRW metric.

Supported by the University of \L{}ód\'{z} grant.

\bibliographystyle{elsarticle-harv}
\bibliography{Perturbative_quantum_damping_of_cosmological_expansion}

\end{document}